\documentstyle[preprint,aps,epsf]{revtex}
\begin{document}
\frenchspacing
\draft
\title{Frustration - how it can be measured}
\author{S. Kobe and T. Klotz}
\address{Technische Universit\"{a}t Dresden,
Institut f\"{u}r Theoretische Physik, D-01062 Dresden, Germany}
\date{20 March 1995}
\maketitle

\begin{abstract}
A misfit parameter is used to characterize the degree of frustration
of ordered and disordered systems. It measures the increase
of the ground-state energy due to frustration in comparison with that
of a relevant reference state. The misfit parameter is calculated
for various spin-glass models. It allows one to compare these models with
each other. The extension of this concept to other combinatorial
optimization problems with frustration, e.g.
$p$-state Potts glasses, graph-partitioning problems and coloring
problems is given. \\
\end{abstract}
\narrowtext

It is well established that two ingredients are necessary to characterize
a spin glass: frustration (competition among the different interactions
acting on a certain magnetic moment) and disorder, see \cite{binyoung}
for reviews. But up to now a quantitative description of
frustration seems to be incomplete. Toulouse \cite{toul77} has introduced
the function $\Phi  =\prod _{(c)} I_{ij}$
 which measures the frustration effect in
a local region of a lattice, where c indicates a closed contour along
the $I_{ij} =+I$  or $-I$ bonds. However, this function cannot be simply
generalized
to other spin-glass models, especially, it is not suited to models without
underlying lattices.\\
Frustration has an effect on ground-state energy and entropy. This can be
easily seen starting from a (unfrustrated) ferromagnetic system by
replacing $+I$ bonds by $-I$ bonds with increasing concentration $p$
\cite{bend94}.
The ground-state energy increases up to a critical concentration $p_c$.
Near $p_c$ the ground-state entropy starts to increase.
This is reflecting the fact that the problem to find
the ground state becomes a problem of combinatorical optimization
with a large number of optimal and nearly optimal solutions. The aim of
this letter is to use just the energy increase due to frustration as
its global measure.\\
Firstly this concept was used to characterize the frustration effect
in an amorphous Ising model with antiferromagnetic short-range interactions
\cite{kobe}. A misfit parameter

\begin{equation}
                  m = \frac{|E^{id}| - |E_0|}{|E^{id}|}
\end{equation}
was introduced, where $E_0$ is the ground-state energy of the frustrated
system and $E^{id}$ is the ground-state energy of a relevant unfrustrated
reference system. The latter can be obtained by replacing all negative
bonds by positive ones. For the $\pm I$ spin glass the relation to
Toulouse's frustration function can be seen by the expression
given by Barahona \cite{barah82}

\begin{equation}
                \begin{array}{cccc} E_0 = -\sum  &|I_{ij}|+2  &
\sum & |I_{ij}| \; , \\[-0.8ex]
\hspace*{3em}{\mbox{\footnotesize $<ij>$}} &  & \mbox{\footnotesize
unsatisfied} & \\[-1.2ex]
& & \mbox{\footnotesize edges} &  \end{array}
\end{equation}
where the first term represents $E^{id}$ and the second one
the numerator of (1) having in mind that the restricted sum over
unsatisfied edges is correlated to Toulouse's function by the total
string length at minimal matching of elementary plaquettes with $\Phi = - 1$.\\
The misfit parameter $m$ of Eq. (1) is used to characterize
the frustration in Ising zig-zag chains in dependence on the chain
length \cite{ziel84} and in a neural network model \cite{schuett89}.
It has been generalized to
quantum systems \cite{rikowo80} \cite{riko82} and to define
the local misfit and the misfit of a cluster of spins
\cite{riko82}.\\
Now we introduce a modified misfit parameter. For a given state $i$ of a
system it is defined by
\begin{equation}
\mu_i = \mu (E_i) = \frac{E_i - E^{id}_{min}}{E^{id}
_{max} - E^{id}_{min}} \; ,
\end{equation}
where $E_i$ is the energy of the state $i$.
$E^{id}_{min}$ and $E^{id}_{max}$ describe the minimal and
maximal ideally possible energy values, respectively,
where 'ideal' refers to the assumption that
all local energies yield a minimal (maximal) contribution to the total
energy. For spin glasses these energies have
to be calculated assuming that all bonds are satisfied
(nonsatisfied). Although, in general, $E^{id}_{min}$ and $E^{id}_{max}$
do not represent necessarily energies of a real system, they often can be
identified with energies of a special reference system.
In any case, they represent lower and upper bounds for
the possible energy range of the considered frustrated system.
Therefore $\mu_i$ is restricted to the interval between $0$ and
$1$. We define the misfit parameter for a
system by the misfit of its ground state
${\mu}_0 = \mu (E_0)$.\\
To clarify the term 'ideal' energy, we will discuss the
misfit parameter of a spin glass. As mentioned above,
the minimal and maximal ideal energies correspond to a
fictive state, where all interactions are satisfied and
nonsatisfied, respectively:
\begin{equation}
E^{id}_{min} = - E^{id}_{max} = -\sum _{<ij>}
|I_{ij}| \; ,
\end{equation}
where the sum goes over all interactions.
Obviously, $E^{id}_{min}$ is the ground state energy of an unfrustrated
reference system, which can be obtained by replacing all $I_{ij}$
by their absolute values. The misfit parameter can be calculated
from Eqs. (3) and (4) as
\begin{equation}
\mu_0 = \frac{1}{2} \left ( 1+ \frac{E_0}{\sum _{<ij>}|I_{ij}|}
\right ) \; .
\end{equation}
Compared to Eq. (1) we get $m = 2 {\mu}_0$. For the $\pm I$
spin glass $\mu_0$ is the fraction of nonsatisfied bonds in the ground state
\cite{stein}.\\
For the well-known
Sherrington-Kirkpatrick (SK) model \cite{SK} the minimal ideal
energy belongs to the ground-state energy of a
reference system, in which the probability
distribution of the interactions is a Gaussian one folded
about zero \cite{chakra} In this case Eq. (4 ) gives
\begin{equation}
E^{id}_{min}/b = - E^{id}_{max}/b = - \sqrt{2/\pi} \approx
-0.798,
\end{equation}
where $b$ denotes the total number of bonds in the
system. Eq. (6) leads to ${\mu}_0 = 0$ for the
mean-field solution in \cite{SK}. In other words, due to
the mean-field approximation the frustration in the
system vanishes and the resulting system is a Mattis-like
spin glass.
The misfit value $\mu_0$  for Parisi's improved replica solution \cite{parisi}
is given in Table I together with a collection of
data for various spin-glass models and related combinatorial
optimization problems.\\
Derrida \cite{derr} has considered the random-energy model as
an approximation to spin-glass models and has calculated lower
bounds for the ground-state energies in any dimension.
For the $\pm I$ spin glass on a $d$-dimensional  hypercubic lattice
this approximation yields reasonable misfit values
($\mu_0 \ge 0.11$ and $0.1$7 for $d = 2$ and $3$, respectively,
and $\mu_0 \ge 0.5 - 1/\sqrt{2d/ln2}$ in the high-dimension limit
$d \rightarrow \infty$, which is lower than the lower limit of the fully
frustrated $\pm I$ system (see Table I)). Otherwise, for the symmetric
Gaussian model on a square lattice the ground-state
energies of the random-energy model are lower than $E^{id}_{min}$.\\

\mediumtext
\begin{table}
\caption{Misfit parameter $\mu_0$ for various spin-glass models and other
combinatorical optimization problems in the ground state.}
\begin{tabular}{|l|l|ll|l|}
\hline
Model & ${\mu}_0$ & \multicolumn{2}{l|}{Remark (Size/Method)} & Ref. \\
\hline
Mattis SG & 0 & & & \cite{mattis}\\
\hline
SK model       & 0               & & mean field & \cite{SK} \\
               & 0.0217 & & replica solution & \cite{parisi} \\
\hline
2d Gaussian SG & $0.090$ & square 16x16 & transfer matrix & \cite{morgen} \\
                & 0.0868 & square 30x30 & exact & \cite{gr} \\
\hline
3d Gaussian SG & $0.143$ & cubic 4x4x4 & transfer matrix & \cite{morgen}\\
               & $0.160$ & cubic 10x10x10 & projected  & \cite{canisius} \\
               & & & gradient method & \\
\hline
2d $\pm I$ SG & $ 0.09$ & honeycomb 12x5 & exact & \cite{lebrecht} \\
          & 0.15 & square 8x8      & exact & \cite{vogel} \\
              & 0.1515 & square 48x48  & multicanonical & \cite{celik} \\
              & 0.14975 & square$^{a}$ & genetic & \cite{gop}\\
              & 0.22 & triangular 6x6  & exact & \cite{vogel} \\
\hline
3d $\pm I$ SG   & 0.211 & cubic 4x4x4 & exact & \cite{klotz/kobe} \\
               & 0.201 & cubic 12x12x12 & multicanonical & \cite{berg}      \\
               & 0.20233 & cubic$^{a}$ & genetic & \cite{gop}\\
\hline
fully frustrated & & & & \\
$\pm I$ systems & 0.25  & $d$-dim. hypercubic & exact & \cite{villain}
\cite{derrida} \\
          & &  $d = 2,3,4$ & & \\
          & 0.3125  & $6d$ hypercubic & exact & \cite{derrida} \\
          & $0.5 - 1/(2\sqrt{d})$ & $d \geq 8$ hypercubic & lower limit &
\cite{derrida} \\
          & $0.5 - 1/\sqrt{2d\pi}$ & $d \geq 8$ hypercubic & upper limit &
\cite{derrida} \\
          & 0.333   & $2d$ triang. / $3d$ fcc & exact & \cite{wannier}
\cite{pincus}\\
          & 0.417 & $6d$ fcc & exact & \cite{pincus} \\
          & $0.5 - 1/(2d)$ & $d$-dim. fcc  & exact & \cite{pincus} \\
\hline
$p$-partitioning ($\pm I$)& & Bethe lattice & analytical approach & \cite{oliv}
\\
$p = 2^{b}$        & 0.074 &  $z=3$ &   &  \\
               & 0.230 &  $z=8$ &  &                  \\
               & 0.257 &  $z=10$ &  &                  \\
$p = 3$        & 0.105 &  $z=3$ &  &  \\
               & 0.317 &  $z=8$ & &                  \\
               & 0.354 &  $z=10$ & &                  \\
\hline
$p$-coloring ($\pm I$)   & & Bethe lattice & analytical approach &
\cite{oliv}\\
$p = 3$        & 0 &  $z=3$ &  &   \\
               & 0.032 &  $z=8$ & &                  \\
               & 0.058 &  $z=10$ & &                  \\
\hline
$p$-state $\pm I$ Potts glass & & Bethe lattice & & \\
$p = 2$        & 0.080 &  $z=3$ & MC and annealing & \cite{banavar}  \\
               & 0.134 &  $z=4$ & & \\
               & 0.1975 &  $z=6$ & & \\
               & 0.236 &  $z=8$ & &                  \\
               & 0.265 &  $z=10$ & &                  \\
$p = 3$        & 0.0068 &  $z=3$ & analytical approach & \cite{oliv}  \\
               & 0.172 &  $z=8$ & &                  \\
               & 0.204 &  $z=10$ & &                  \\

\end{tabular}
$^{\rm a}$extrapolated from $1/N$ scaling\\
$^{\rm b}$The misfit for the corresponding ($p=2$)-coloring
        problem results in the same $\mu_0$ values.
\end{table}
\newpage
\narrowtext
{}From Table I the various effects of
dimension, coordination number,
distribution and range of interactions and number of states
per spin variable on the frustration can be seen in a quantitative manner. \\
At least for small coordination numbers and dimensions
their increases result in additional constraints and therefore
in increasing $\mu_0$ values. The effect of different coordination
numbers can be seen both for the two-dimensional $\pm I$ Ising
spin glasses with different lattice structures and for the $p$-state
$\pm I$ Potts glass, the $p$-partitioning and the $p$-coloring problem with
different $z$ values. The influence of different dimensions can be
studied by comparing the results for two and three dimensions.
The comparison between two- and three-dimensional $\pm I$
spin glasses and spin glasses on a Bethe lattice with the same
number of nearest neighbors $z$ shows that $\mu_0$ is stronger
influenced by the coordination number than by the spatial structure
and dimension. However, it can be seen also that $\mu_0$ is
lower in the Bethe lattice than in higher netted lattices.\\
Analytical expressions for $\mu_0(d)$ are given for hypercubic
fully frustrated systems \cite{derrida} and $d$-dimensional antiferromagnets
with triangular plaquettes \cite{pincus} at least in the
high-dimension limit. For finite $d$ the results for fully
frustrated systems are proved as upper bounds for systems
with equal probability of $+ I$ and $- I$ bonds.\\
In systems with a Gaussian distribution of interactions the energy
can be decreased by choosing and frustrating that bond with the
lowest strength in a plaquette. Therefore the misfit for such systems
is smaller than for comparable systems with a $\pm I$ distribution.\\
Other relations between the parameters of a model and the
resulting frustration can be investigated by using the $p$-state Potts
glass, the $p$-partitioning and the $p$-coloring problem. With an
increasing number of colors $p$ for the nearest neighbours of a
site in a $p$-coloring problem the chance increases to give all neighbours
another colors and therefore the frustration in the
system decreases. On the other hand, if the number $p$ of
subsets in the $p$-partitioning problem increases, the problem
becomes more complicated and restricted, leading to an
increase in frustration. In Potts glasses the number of possibilities
to avoid frustration increases with rising number $p$ of
states per spin and consequently $\mu_0$ decreases.
As outlined in \cite{oliv} the $p$-state Potts glass can be
understood approximately as an intermediate system between
the $p$-partitioning and the $p$-coloring problem, which correspond
to a ferromagnetic and an antiferromagnetic Potts glass with
special magnetization constraints. Therefore, for the same $p$
the $\mu_0$ values in the Potts glass are smaller than those of
the partitioning problems but larger than those of
the coloring problems. Small deviations for the case $p = 2$
are due to different methods.\\
Summarizing the results, we have introduced a global misfit parameter
generalizing
the fraction of unsatisfied bonds in the ground state of the
short-range $\pm I$ spin glass
to other spin glass models and to other systems with frustration.
It measures the influence of frustration on the ground-state energy
and, for the first time, it allows to compare
various spin glass models quantitatively. It can
be applied to systems without an underlying lattice structure.
An advantage over the
former parameter (1) is its invariance against any linear scaling of
energies. It means, e.g., that additional self energy terms leave
the misfit unchanged. \\
The presented concept can be applied to other systems with frustration
at least for such problems, for which the cost functions can be transformed
linearly without additional constraints. This can be understood in terms of
the existing transformations between various problems using the
NP-completeness. But often, e.g. for the traveling salesman problem,
such a transformation is accompanied with new global constraints.
We will focus on this topic in a forthcoming paper.\\
The presented parameter only refers to the {\it energetic} aspect of
frustration. It is an open question whether a similar parameter
can be found for the {\it entropic} characterization of frustrated
systems. A preliminary answer for the $\pm I$ models is given by
Vogel et al. \cite{vogel}. These authors have calculated the fraction of
bonds, which are satisfied in {\it all} ground states. The difference
between unity and this fraction can be used
as a global entropic measure for frustration. Generalizations are
under consideration.\\
We benefit from discussions with A.R. Ferchmin, A. Hartwig, K.-H. Hoffmann,
A. M\"obius, P. Polaszek, H. Rieger, E. E. Vogel
and J. Wei{\ss}barth. This work is supported by the DFG
(project no. Ko 1416).


\begin{references}
\bibitem[1]{binyoung}K. Binder and A. P. Young, Rev.\ Mod.\ Phys. {\bf 58},
           801 (1986);
           M. M\'ezard, G. Parisi, and M. A. Virasoro, {\it
           Spin Glass Theory and Beyond} (World Scientific, Singapore, 1987);
           K. Fischer and J. A. Hertz, {\it Spin Glasses}
           (Cambridge University Press, Cambridge, 1991).
\bibitem[2]{toul77}G. Toulouse, Commun.\ Phys. {\bf 2}, 99 (1977).
\bibitem[3]{bend94}S. Kirkpatrick, Phys.\ Rev.\ B  {\bf 16}, 4630 (1977);
        M. Achilles, J. Bendisch, K. Cassirer, H. von Trotha,
        GMD-Studien Nr. 186 (Gesellschaft f\"ur Mathematik und
        Datenverarbeitung mbH, Sankt Augustin, Germany, 1991).
\bibitem[4]{kobe}S. Kobe and K. Handrich, phys.\ stat.\ sol.\ (b) {\bf 73},
        K65 (1976); S. Kobe, in {\it Amorphous Magnetism II}, ed. by R. A.
        Levy and R. Hasegawa (Plenum, New York, 1977), p. 529.
\bibitem[5]{barah82}F. Barahona, J.\ Phys.\ A:\ Math.\ Gen. {\bf 15}, 3241
(1982).
\bibitem[6]{ziel84}W. Zieli\'nski, A. R. Ferchmin, and S. Kobe,
        Phys.\ Lett. {\bf 102A}, 66 (1984).
\bibitem[7]{schuett89}S. Kobe and A. Sch\"utte, Acta\ Phys.\ Polon.
        {\bf A75}, 891 (1989).
\bibitem[8]{rikowo80}J. Richter, S. Kobe, and H. Wonn,
        phys.\ stat.\ sol. (b) {\bf 98}, K37 (1980).
\bibitem[9]{riko82}J. Richter and S. Kobe, J.\ Phys.\ C:\ Solid\ State\
        Phys. {\bf 15}, 2193 (1982).
\bibitem[10]{stein}D. L. Stein, G. Baskaran, S. Liang, and M. N. Barber,
        Phys.\ Rev.\ B\ {\bf 36}, 5567 (1987).
\bibitem[11]{SK}D. Sherrington and S. Kirkpatrick, Phys.\ Rev.\ Lett.\
        {\bf 35}, 1792 (1975); S. Kirkpatrick and D. Sherrington, Phys.\
        Rev.\ B\ {\bf 17}, 4384 (1978).
\bibitem[12]{chakra}A. Chakrabarti and R. Toral, Phys.\ Rev.\ B\ {\bf 39},
        542 (1989).
\bibitem[13]{parisi}G. Parisi, J.\ Phys.\ A:\ Math.\ Gen.\ {\bf 13}, L115
        (1980).
\bibitem[14]{derr}B. Derrida, Phys.\ Rev.\ B\ {\bf 24}, 2613 (1981).
\bibitem[15]{mattis}D. C. Mattis, Phys.\ Lett.\ A\ {\bf 56}, 421 (1976).
\bibitem[16]{morgen}I. Morgenstern and K. Binder, Phys.\ Rev.\ B\
        {\bf 22}, 288 (1980); Z.\ Phys.\ B\ {\bf 39}, 227 (1980).
\bibitem[17]{gr}M. Gr\"otschel, M. J\"unger, and G. Reinelt, in {\it Heidelberg
Colloquium
        on Glassy Dynamics}, Lecture Notes in Physics 275, edited by
        J. L. van Hemmen and I. Morgenstern (Springer-Verlag, Berlin, 1987), p.
325.
\bibitem[18]{canisius}J. Canisius and J. L. van Hemmen, Europhys.\ Lett.\
        {\bf 1}, 319 (1986).
\bibitem[19]{lebrecht}W. Lebrecht and E. E. Vogel, private communication.
\bibitem[20]{vogel} E. E. Vogel, J. Cartes, S. Contreras, W. Lebrecht, and
            J. Villegas, Phys.\ Rev.\ B\ {\bf 49}, 6018 (1994).
\bibitem[21]{celik} T. Celik, U. H. E. Hansmann, and B. Berg, in {\it Computer
        Simulation Studies in Condensed-Matter Physics VI}, Springer Proc. in
Physics
        Vol. 76, edited by D. P. Landau, K. K. Mon, and H.-B. Sch\"uttler
(Springer-Verlag,
        Berlin, 1993), p. 173.
\bibitem[22]{gop}U. Gropengiesser, Physica\ A, in the press.
\bibitem[23]{klotz/kobe}T. Klotz and S. Kobe, J.\ Phys.\ A:\ Math.\ Gen.\
        {\bf 27}, L95 (1994).
\bibitem[24]{berg}B. A. Berg, U. E. Hansmann, and T. Celik, Phys.\ Rev.\ B\
        {\bf 50}, 16444 (1994).
\bibitem[25]{villain}J. Villain, J.\ Phys.\ C:\ Solid\ State\ Phys.\ {\bf10},
        L537 (1977).
\bibitem[26]{derrida}B. Derrida, Y.Pomeau, G. Toulouse, and J. Vannimenus,
        J.\ Phys.\ (Paris)\ {\bf 40}, 617 (1979).
\bibitem[27]{wannier}G. H. Wannier, Phys.\ Rev. {\bf 78}, 341 (1950).
\bibitem[28]{pincus}S. Alexander and P. Pincus, J.\ Phys.\ A:\ Math.\ Gen.\
        {\bf 13}, 263 (1980).
\bibitem[29]{oliv}M. J. de Olivera, J.\ Stat.\ Phys.\ {\bf 54}, 477 (1989).
\bibitem[30]{banavar}J. R. Banavar, D. Sherrington, and N. Sourlas,
        J.\ Phys.\ A:\ Math.\ Gen.\ {\bf 20}, L1 (1987).
\end{references}
\end{document}